\documentclass[journal,twoside]{IEEEtranTCOM}

\usepackage{cite,bm,tikz,amsfonts,amssymb,graphicx,pgfplots,multirow,epstopdf,algorithm,algorithmicx,algpseudocode}
\usepackage[cmex10]{amsmath}
\usetikzlibrary{arrows.meta,decorations.pathreplacing,plotmarks,shapes,arrows,positioning}
\pgfplotsset{compat=1.12}
\pgfplotsset{minor grid style={dotted,white!70!black}}
\pgfplotsset{every tick label/.append style={font=\scriptsize}}
\let\originalleft\left
\let\originalright\right
\renewcommand{\left}{\mathopen{}\mathclose\bgroup\originalleft}
\renewcommand{\right}{\aftergroup\egroup\originalright}

\newcommand{\define}{\triangleq}
\newcommand{\var}[1]{\mathsf{Var}\left(#1\right)}

\newcommand{\EVi}[1]{\mathsf{E}\left[#1\right]}
\newcommand{\EVe}[2]{\mathsf{E}_{#1}\left[#2\right]}
\newcommand{\modset}{\mathcal X}
\newcommand{\varLPN}{\sigma_\phi^2}
\newcommand{\varDrift}{\sigma_\delta^2}
\newcommand{\varN}[1]{\sigma_{#1}^2}
\newcommand{\varNtilde}[1]{\bar\sigma_{#1,k}^2}
\newcommand{\snr}{\frac{E_\mathrm{b}}{N_0}}
\newcommand{\snrinline}{E_\mathrm{b}/N_0}
\newcommand{\Nit}{T}
\newcommand{\cgaussian}[3]{\mathcal{CN}_{#1}\left(#2,#3\right)}
\newcommand{\gaussian}[3]{\mathcal N_{#1}\left(#2,#3\right)}
\newcommand{\gaussianInline}[3]{\mathcal N_{#1}(#2,#3)}
\newcommand{\tikhonov}[2]{\mathcal T_{#1}\left(#2\right)}
\newcommand{\qGamma}[3]{q_{#1}^{(#2)}(#3)}
\newcommand{\qGammaSI}[2]{q_{#1}(#2)}
\newcommand{\qGammai}[2]{q_{#1}^{(#2)}}
\newcommand{\qGammaiSI}[1]{q_{#1}}
\newcommand{\qS}[3]{q_{#1}^{(#2)}(#3)}
\newcommand{\qSSI}[2]{q_{#1}(#2)}
\newcommand{\qSi}[2]{q_{#1}^{(#2)}}
\newcommand{\qSiSI}[1]{q_{#1}}
\newcommand{\Es}{E_\mathrm{s}}
\newcommand{\Pd}[2]{P_\mathrm{d}(#2)}
\newcommand{\Pu}[2]{P_\mathrm{u}(#2)}

\newcommand{\pd}[2]{p_\mathrm{d}(#2)}

\DeclareMathOperator*{\argmax}{argmax}
\DeclareMathOperator*{\diag}{diag}

\begin{document}
	
	\title{Iterative Detection and Phase-Noise Compensation for Coded Multichannel Optical Transmission}
	
	\author{Arni~F.~Alfredsson,~\IEEEmembership{Student~Member,~IEEE},~Erik~Agrell,~\IEEEmembership{Fellow,~IEEE},~and~Henk~Wymeersch,~\IEEEmembership{Member,~IEEE}
		\thanks{A. F. Alfredsson, E. Agrell, and H. Wymeersch are with the Department of Electrical Engineering, Chalmers University of Technology, SE-41296 G\"{o}teborg, Sweden (e-mail: arnia@chalmers.se; agrell@chalmers.se; henkw@chalmers.se).
			
			This work was supported by the Swedish Research Council (VR) under Grants 2013-5642 and 2014-6138.
			
			%	Copyright \copyright{} 2019 IEEE. Personal use of this material is permitted. However, permission to use this material for any other purposes must be obtained from the IEEE by sending a request to pubs-permissions@ieee.org.
	}}
	\maketitle
	
	\begin{abstract}
		The problem of phase-noise compensation for correlated phase noise in coded multichannel optical transmission is investigated. To that end, a simple multichannel phase-noise model is considered and the maximum \textit{a posteriori} detector for this model is approximated using two frameworks, namely factor graphs (FGs) combined with the sum--product algorithm (SPA), and a variational Bayesian (VB) inference method. The resulting pilot-aided algorithms perform iterative phase-noise compensation in cooperation with a decoder, using extended Kalman smoothing to estimate the \textit{a posteriori} phase-noise distribution jointly for all channels. The system model and the proposed algorithms are verified using experimental data obtained from space-division multiplexed multicore-fiber transmission. Through Monte Carlo simulations, the algorithms are further evaluated in terms of phase-noise tolerance for coded transmission. It is observed that they significantly outperform the conventional approach to phase-noise compensation in the optical literature. Moreover, the FG/SPA framework performs similarly or better than the VB framework in terms of phase-noise tolerance of the resulting algorithms, for a slightly higher computational complexity. 
	\end{abstract}
	
	
	\section{Introduction}
	\label{sec:intro}
	
	Phase noise is an inherent problem in optical communications. This comes due to the nonzero linewidth of light sources and local oscillators (LOs) \cite{Ip:08}, which degrades the system performance severely if not handled properly. This is particularly relevant since the research focus has shifted in recent years towards higher-order quadrature amplitude modulation (QAM) or more advanced multilevel modulation formats \cite{Karlsson:17}. In general, systems become more sensitive to phase noise as the modulation order grows, and hence, effective phase-noise compensation becomes crucial. Traditionally, phase-noise compensation methods in optical communications have been blind, i.e., they do not use pilot symbols to assist with the estimation, and thus, spectral efficiency is not sacrificed. Well-known examples are the Viterbi--Viterbi algorithm \cite{1056713} and blind phase search (BPS) \cite{4814758}. However, blind methods suffer from ambiguity in the phase-noise estimation and are therefore susceptible to cycle slips, which result in burst errors \cite{4298982} that can not be corrected with a code. This can be handled with differential encoding, which has the downside of increasing the average bit error rate (BER). Alternatively, one can resort to pilot-aided phase-noise estimation \cite{1504908,4799047,6855336}.
	
	Recently, space-division multiplexing (SDM) has been a topic of particular interest. It involves the integration of various system components, such as optical hardware and digital signal processing (DSP) algorithms, as well as the utilization of multicore and multimode fibers, combinations thereof, or bundles of single-mode fibers \cite{sdm:kahn17}. These systems enable the joint-channel compensation of various signal impairments, in particular phase noise, as light sources and LOs can be shared between different cores or modes \cite{6317137}. This gives rise to spatial correlation in the phase noise, which can be exploited to relax hardware requirements \cite{alfredsson2017ofc} and reduce receiver complexity \cite{7121584}. However, the phase noise is not perfectly correlated, as environmental effects and system imperfections introduce phase drifts between cores and polarizations \cite{7183869,6317137}. In addition, correlated phase noise is present in wavelength-division multiplexed (WDM) systems utilizing frequency combs, as the combs enable phase locking between the different frequency lines. However, similarly to SDM systems, the phase noise is not perfectly correlated among the spectral channels, due to imperfections in the comb generation \cite{LarsECOC:17}. Joint-channel phase-noise compensation has also been demonstrated in systems utilizing electrically generated subcarriers \cite{6290331}. Clearly, various types of optical systems are amenable to joint-channel phase-noise compensation, which is the target of this investigation.
	
	Multiple solutions that exploit the phase-noise correlation between channels have been proposed for SDM and comb-based WDM systems. The majority has focused on schemes that facilitate DSP complexity reduction\cite{6317137,6517220,LarsECOC:17,6290331}, e.g., through the use of a master--slave strategy, in which one channel is used to produce phase-noise estimates that are used for compensation in all channels. However, the phase-noise correlation can also be used to improve performance in terms of increased phase-noise tolerance \cite{8013014} or, in the case of pilot-aided schemes, lower the pilot rate while maintaining the same phase-noise tolerance.
	
	Wireless communication systems face a similar problem with oscillator phase noise, which has given rise to a myriad of solutions that approximate the maximum \textit{a posteriori} (MAP) detector for phase-noise channels. In particular, multiple algorithms that perform iterative phase-noise compensation and data detection have been proposed. Pertaining to single-channel transmission, blind and pilot-aided iterative solutions have been developed based on factor graphs (FGs) and the sum--product algorithm (SPA) \cite{1504908,7130553,6855336,1228245,7349155}, variational Bayesian (VB) inference \cite{4799047}, and the expectation-maximization algorithm \cite{Noels:2005:TFS:1086710.1086714,5069546}. Moreover, in the context of multiple-input multiple-output (MIMO) systems for wireless transmission, various methods have been proposed for joint-channel phase-noise compensation \cite{1400262,6655039,8253485}. In \cite{7080901,6809377}, several algorithms were proposed for joint phase-noise estimation and data detection using the aforementioned frameworks, and in \cite{6212402,6422412}, joint channel and phase-noise estimation for MIMO systems was proposed. The majority of the work has focused on multichannel models entailing oscillator phase noise that is either identical or independent between antennas in the MIMO system, in addition to channel mixing.
	
	In this paper, we consider coded multichannel optical transmission in the presence of correlated phase noise and propose algorithms for iterative phase-noise estimation and data detection, which has yet to be addressed in the context of optical communications. The contributions can be summarized as follows:
	(i) We consider a simple multichannel system model that entails arbitrarily correlated phase noise and approximates the received signal after it has been processed by a typical DSP chain prior to phase-noise compensation, which has not been considered for certain multichannel optical transmission scenarios such as multicore-fiber systems, with the purpose of facilitating algorithm development for joint detection and phase-noise compensation. Using this model, we use two frameworks that have proven to be effective for similar system models in wireless communications \cite{1504908,4799047,7080901}, namely FG/SPA\cite{910572} and VB inference \cite{beal:varbayes}, to develop algorithms that perform iterative detection and phase-noise estimation in fiber-optical transmission. These frameworks yield different interfaces between the phase-noise estimation and the decoding. A trade-off in terms of performance and computational complexity is observed between the resulting algorithms;
	(ii) The phase-noise estimation consists of the extended Kalman filter (EKF) and the Rauch--Tung--Striebel smoother (RTSS). We show that the standard EKF equations can be simplified for the system model, and that the linearization imposed by the EKF on the system model does not penalize the phase-noise estimation performance for practical baud rates and laser linewidths;
	(iii) Finally, we demonstrate the effectiveness of the system model and the proposed algorithms using experimental data from SDM transmission through a weakly-coupled, homogeneous, multicore fiber. We then further evaluate the algorithms through Monte Carlo simulations of coded transmission in terms of phase-noise tolerance, and show that they significantly outperform the approach that is typically used in the optical literature: Per-channel phase-noise compensation using BPS and symbol detection based on minimum Euclidean distance.
	
	\textit{Notation:} The real part, imaginary part, complex conjugate, and angle of a complex number are denoted with $\Re\{\cdot\}$, $\Im\{\cdot\}$, $(\cdot)^*$, and $\angle(\cdot)$, resp. Random variables and their realizations are denoted with $X$ and $x$, resp. The expectation of a random variable $X$ with respect to a distribution $P$ is written as $\EVe{P}{X}$. Probability mass functions (PMFs) are written as $P(x)$, whereas probability density functions (PDFs) and mixed discrete--continuous distributions are denoted with $p(x)$. In particular, a multivariate real Gaussian PDF with mean $\bm\mu$, covariance matrix $\mathbf\Sigma$, and argument $\bm x$ is denoted with $\gaussian{\bm x}{\bm\mu}{\mathbf\Sigma}$, and its complex counterpart with argument $\bm z$ is written as $\cgaussian{\bm z}{\bm\mu}{\mathbf\Sigma}$.	Scalars, vectors, and matrices are typeset as $x$, $\bm x$, and $\mathbf X$, resp. A diagonal matrix is denoted with $\diag(\cdot)$, whereas the identity matrix of size $D$ is written as $\mathbf I_D$. Finally, the vector transpose is denoted with $(\cdot)^T$.
	
	\section{System Model}
	\label{sec:sysmodel}
	Transmission over $D$ parallel complex-valued channels is considered. The transmitted frame in each channel is modeled as a vector of $N$ random variables that take on values from a set $\modset$ of constellation points. The constellation is normalized such that the mean of the constellation points is zero and the average symbol energy is $\Es$. The received signal is assumed to have undergone standard DSP steps such as resampling, electronic dispersion compensation, orthonormalization, timing recovery, adaptive equalization, and frequency-offset estimation. Furthermore, the adaptive equalization is assumed to have been carried out in a phase-immune fashion using, e.g., a radially-directed equalizer \cite{4909145}. With these assumptions in place, the received signal in each channel, after further resampling to one sample per symbol, is approximated as the output of a system entailing additive white Gaussian noise (AWGN) and phase noise. In single-channel optical systems, this is also a typical assumption of the processed signal prior to phase-noise compensation \cite{4814758,4298982,5419121}. The discrete-time complex baseband model is therefore written as
	\begin{equation}
		r_{i,k} = s_{i,k}e^{j\theta_{i,k}}+n_{i,k},\label{eq:sysmodel}
	\end{equation}
	where $k=1,\dots,N$ is a time index and $i=1,\dots,D$ is a channel index. The received samples, transmitted symbols, phase noise, and complex AWGN realizations are denoted with $r_{i,k}$, $s_{i,k}$, $\theta_{i,k}$, and $n_{i,k}$, resp., where the complex AWGN on channel $i$ has variance $\varN{i}$ per real dimension. The vector $\bm r_k=[r_{1,k},\dots,r_{D,k}]^T\in\mathbb C^{D}$ contains the received samples in all channels at time $k$, and the vectors $\bm s_k$, $\bm \theta_k$, and $\bm n_k$ are defined analogously. Finally, let $\bm r$ contain all received samples, and $\bm s$, $\bm\theta$, and $\bm n$ be defined similarly.
	
	It has been experimentally demonstrated that for optical multichannel transmission through, e.g., optical frequency combs \cite{zanette2017correlation} or multicore fibers \cite{6317137} with shared lasers, the phase noise is highly correlated among the channels, albeit not fully owing to per-channel phase drifts. These drifts are caused by environmental effects as well as imperfections in hardware and DSP and are much slower than the laser phase noise, which is typically modeled as a Gaussian random walk. To describe the laser phase noise in addition to slower channel-specific drifts, we opt for a simplistic phase-noise model to facilitate the development of high-performance algorithms based on detection and estimation theory\footnote{Whenever simplistic models are used for algorithm design, it is imperative to assess the efficacy of the model and the resulting algorithms with either more realistic models or experimental data. In Sec.~\ref{sec:exp_ver}, the latter approach is taken.}. Thus, the phase noise across all channels is approximated with a multidimensional Gaussian random walk, as
	\begin{equation}
		\bm\theta_k=\bm\theta_{k-1}+\Delta{\bm\theta}_k,\label{eq:random_walk}
	\end{equation}
	where $\bm\theta_1$ is uniformly distributed on $[0,2\pi)^{D}$ and $\Delta{\bm\theta}_k$ is a multivariate zero-mean Gaussian random variable with covariance matrix $\mathbf Q$. This covariance matrix depends on the system and has to be known or estimated. Furthermore, the phase noise is assumed to be independent of the transmitted symbols and AWGN, and unknown to the transmitter and receiver. Finally, $\mathbf Q$ and $\bm\sigma^2=[\varN{1},\dots,\varN{D}]$ are assumed known to the receiver.
	
	The transmitted symbol sequence $\bm s$ is assumed to be obtained through an encoding function $f(\cdot)$ that maps an equiprobable information bit sequence $\bm b$ of length $N_\mathrm{b}$ to a data symbol sequence of length $ND$. Additional pilot symbols that are used for the phase-noise compensation are considered as a part of the channel code. The pilot distribution is assumed known to the transmitter and receiver.
	
	\section{Derivation of Algorithms}
	\label{sec:alg}
	
	The MAP detector yields the lowest possible BER out of all detectors \cite[Ch.~1.4]{ryan2009channel}. It performs detection on a bit-by-bit basis according to
	\begin{equation}
		\hat b_l=\argmax_{b_l\in\{0,1\}}P\left(b_l|\,\bm r\right),
		\label{eq:map}
	\end{equation}
	for $l=1,\dots,N_\mathrm{b}$, where $P(b_l|\bm r)$ is the \textit{a posteriori} PMF of $b_l$. However, the PMF in (\ref{eq:map}) is hard to compute exactly for the system model in (\ref{eq:sysmodel}), and thus, approximations are needed. To that end, note that the desired PMF in (\ref{eq:map}) can be obtained through the marginalization
	\begin{equation}
		P(b_l|\bm r)=\int_{\mathbb R^{D\times N}}\sum_{\bm s\in\modset^{D\times N}}\sum_{\bm b\in\mathcal U_l(b_l)}p(\bm b,\bm s,\bm\theta|\bm r)\mathrm d\bm\theta,\label{eq:marg}
	\end{equation}
	where $\mathcal U_l(b_l)=\left\{ \bm b'\in\{0,1\}^{N_\mathrm{b}}:b_l'=b_l\right\}$. To carry out this marginalization approximately but efficiently, we make use of two frameworks, namely FG/SPA and VB inference, motivated by the fact that they have been found effective in earlier work pertaining to phase-noise compensation in wireless applications \cite{1504908,4799047,7080901,6855336}. Conceptually, the resulting receiver algorithms comprise three blocks: (i) A standard iterative decoder, which has as input probabilities on bits (assuming the use of a binary code); (ii) A scheme that estimates the phase-noise distribution and has as input symbol probabilities; (iii) An interface, developed using the aformentioned frameworks, that connects the decoder and the phase-noise estimation scheme.
	
	As the AWGN is assumed to be independent between channels, moving from single-channel to multichannel transmission does not complicate the computation of bit-wise probabilities that are fed to the decoder, given that the phase-noise distribution is known at each symbol. However, assuming the phase noise to be arbitrarily correlated across the channels in multichannel transmission prevents trivial extensions of various existing single-channel phase-noise estimation methods, such as the well-known Tikhonov algorithm in \cite[Sec.~IV-B]{1504908}. In order to estimate the phase-noise distribution in the multichannel case, one can resort to, e.g., an extended Kalman smoother (EKS), an unscented Kalman smoother, or a particle smoother. In this work, we opt for the EKS, which is naturally suited to estimate the marginal \textit{a posteriori} PDFs at each time $k$ of a multidimensional Gaussian random walk. This approach proved to be successful in wireless communications \cite{6422412}, where one of the proposed algorithms included a soft-input EKS capable of estimating spatially correlated phase-noise processes. In this work, we apply a soft-input EKS in multichannel fiber-optical communications, and accordingly use a somewhat different system model as well as a different method to derive the soft inputs to the EKS compared with \cite{6422412}. The marginal phase-noise distributions are approximated as multivariate Gaussian PDFs by the EKS; hence, they can be further marginalized in a low-complexity manner to yield the phase-noise distribution at each symbol. This also facilitates the derivations of the interfaces connecting the decoder and the phase-noise estimation, as will be shown in the following subsections.
	
	The EKS consists of recursive equations that are used to approximately estimate the marginal \textit{a posteriori} phase-noise PDFs, $p(\bm\theta_k|\bm r)$, for $k=1,\dots,N$. This is accomplished through two recursive passes; a forward pass with an EKF\cite[Ch.~5.2]{sarkka:bayesian} and a backward pass with an RTSS\cite[Ch.~8.2]{sarkka:bayesian}. The resulting recursive equations are used by both frameworks to yield the final algorithms. A further justification for utilizing an EKF is that the linearization imposed by the EKF on the system model in \eqref{eq:sysmodel} works effectively, provided that the phase noise does not vary too rapidly. For practical parameters, this is indeed the case, as illustrated in Section~\ref{sec:linerror}.
	
	\subsection{Phase-Noise Estimation}
	\label{sec:ekf_rtss}
	
	The EKF estimates the marginal PDFs $p(\bm\theta_k|\bm r_1,\dots,\bm r_k)$ for $k=2,\dots,N$. However, instead of applying the standard equations\cite[Ch.~5.2]{sarkka:bayesian} directly, it is shown in Appendix \ref{app:ekf_derivation}\footnote{See also \cite{8253485} for an alternative derivation of the EKF equations for a random-walk phase noise with complex-valued observations in the context of wireless MIMO transmission with independent oscillators.} that they can be reduced to
	\begin{align}
		\mathbf M^{\mathrm f}_{k|k-1}&=\mathbf M^{\mathrm f}_{k-1}+\mathbf Q,\label{eq:ekf_varpred}\\
		\mathbf M^{\mathrm f}_{k}&=(\mathbf I_D+\mathbf M^{\mathrm f}_{k|k-1}\mathbf V_k)^{-1}\mathbf M^{\mathrm f}_{k|k-1},\label{eq:ekf_varcorr}\\
		\bm{\hat\theta}^{\mathrm f}_k&=\bm{\hat\theta}^{\mathrm f}_{k-1}+\mathbf M^{\mathrm f}_{k}\bm h_k,\label{eq:ekf_corr}
	\end{align}
	with initial conditions
	\begin{align}
		\bm{\hat\theta}^{\mathrm f}_{1}&=\left[\angle(r_{1,1}s_{1,1}^*),\dots,\angle(r_{D,1}s_{D,1}^*)\right]^T,\\
		\mathbf M^{\mathrm f}_{1}&=\diag\left(\frac{\varN{1}}{\Es},\dots,\frac{\varN{D}}{\Es}\right),
	\end{align}
	where $\mathbf V_k$ and each element of $\bm h_k=[h_{1,k},\dots,h_{D,k}]^T$ in \eqref{eq:ekf_corr} are computed as
	\begin{align}
		\mathbf V_k&=\diag\left(\frac{|s_{1,k}|^2}{\varN{1}},\dots,\frac{|s_{D,k}|^2}{\varN{D}}\right),\label{eq:S}\\
		h_{i,k}&=\frac{1}{\varN{i}}\Im\left\{r_{i,k}s_{i,k}^*e^{-j\hat\theta^{\mathrm f}_{i,k-1}}\right\}.\label{eq:h}
	\end{align}
	The rationale behind the initialization of $\mathbf M^{\mathrm f}_{1}$ is the fact that for practical signal-to-noise ratios (SNRs), the variance of the elements of $\bm{\hat\theta}^{\mathrm f}_{1}$ will be approximately half the variance of the complex AWGN. Moreover, $\mathbf Q$ is the covariance matrix of $\Delta\bm\theta_k$ in \eqref{eq:random_walk}, whereas $\bm{\hat\theta}^{\mathrm f}_{k-1}=[\hat\theta_{1,k-1}^{\mathrm f},\dots,\hat\theta_{D,k-1}^{\mathrm f}]^T$ denotes an estimate of $\bm\theta_{k-1}$ based on all received samples up to and including time $k-1$. Note that $s_{i,k}$ is assumed to be known for all $i$ and $k$ in \eqref{eq:ekf_varpred}--\eqref{eq:S}. This is not true for $(i,k)$ corresponding to data symbols and will be handled in the following subsections.
	
	To estimate $p(\bm\theta_k|\bm r)=p(\bm\theta_k|\bm r_1,\dots,\bm r_N)$, the RTSS is used. The resulting recursive equations are given by \cite[Ch.~8.2]{sarkka:bayesian}
	\begin{align}
		\mathbf A_k&=\mathbf M^{\mathrm f}_{k}(\mathbf M^{\mathrm f}_{k+1|k})^{-1},\label{eq:gammaA}\\
		\bm{\hat\theta}^{\mathrm s}_k&=\bm{\hat\theta}^{\mathrm f}_{k}+\mathbf A_k(\bm{\hat\theta}^{\mathrm s}_{k+1}-\bm{\hat\theta}^{\mathrm f}_{k}),\label{eq:gammahatmean}\\
		\mathbf M^{\mathrm s}_k&=\mathbf M^{\mathrm f}_{k}+\mathbf A_k(\mathbf M^{\mathrm s}_{k+1}-\mathbf M^{\mathrm f}_{k+1|k})\mathbf A_k^T,\label{eq:gammahatvar}
	\end{align}
	for $k=N-1,N-2,\dots,1$, with initial conditions
	\begin{align}
		\bm{\hat\theta}^{\mathrm s}_N&=\bm{\hat\theta}^{\mathrm f}_{N},\\
		\mathbf M^{\mathrm s}_N&=\mathbf M^{\mathrm f}_{N}.
	\end{align}
	Thus, $\bm{\hat\theta}^{\mathrm s}_k=[\theta_{1,k}^{\mathrm s},\dots,\theta_{D,k}^{\mathrm s}]^T$ and $\mathbf M^{\mathrm s}_k$ represent the mean vector and covariance matrix of the Gaussian approximation of $p(\bm\theta_k|\bm r)$, i.e., $p(\bm\theta_k|\bm r)\approx\gaussianInline{\bm\theta_k}{\bm{\hat\theta}^{\mathrm s}_k}{\mathbf M^{\mathrm s}_k}$. The EKF and RTSS equations are summarized in Algorithm~\ref{alg:ekf_rtss}.
	
	\begin{algorithm}
		\caption{EKS}
		\label{alg:ekf_rtss}
		\begin{algorithmic}[1]
			\Require $\bm r$, $\bm s$, $D$, $N$, $\mathbf Q$, $\bm\sigma^2$
			\Ensure $(\bm{\hat\theta}^{\mathrm s}_k,\mathbf M^{\mathrm s}_k)~\forall~k$
			\State $\bm{\hat\theta}^{\mathrm f}_{1}=\left[\angle(r_{1,1}s_{1,1}^*),\dots,\angle(r_{D,1}s_{D,1}^*)\right]^T$
			\State $\mathbf M^{\mathrm f}_{1}=\diag(\varN{1}/\Es,\dots,\varN{D}/\Es)$
			\For{$k=2,\dots,N$}
			\For{$i=1,\dots,D$}
			\State $h_{i,k}={\Im\{r_{i,k}s_{i,k}^*e^{-j\hat\theta^{\mathrm f}_{i,k-1}}\}}/{\varN{i}}$
			\EndFor
			\State $\mathbf V_k=\diag({\left|s_{1,k}\right|^2}/{\varN{1}},\dots,{\left|s_{D,k}\right|^2}/{\varN{D}})$
			\State $\mathbf M^{\mathrm f}_{k|k-1}=\mathbf M^{\mathrm f}_{k-1}+\mathbf Q$
			\State $\mathbf M^{\mathrm f}_{k}=(\mathbf I_D+\mathbf M^{\mathrm f}_{k|k-1}\mathbf V_k)^{-1}\mathbf M^{\mathrm f}_{k|k-1}$
			\State $\bm{\hat\theta}^{\mathrm f}_{k}=\bm{\hat\theta}^{\mathrm f}_{k-1}+\mathbf M^{\mathrm f}_{k}\bm h_k$
			\EndFor
			\State $(\bm{\hat\theta}^{\mathrm s}_N,\mathbf M^{\mathrm s}_N)=(\bm{\hat\theta}^{\mathrm f}_{N},\mathbf M^{\mathrm f}_{N})$
			\For{$k=N-1,N-2,\dots,1$}
			\State $\mathbf A_k=\mathbf M^{\mathrm f}_{k}(\mathbf M^{\mathrm f}_{k+1|k})^{-1}$
			\State $\bm{\hat\theta}^{\mathrm s}_k=\bm{\hat\theta}^{\mathrm f}_{k}+\mathbf A_k(\bm{\hat\theta}^{\mathrm s}_{k+1}-\bm{\hat\theta}^{\mathrm f}_{k})$
			\State $\mathbf M^{\mathrm s}_k=\mathbf M^{\mathrm f}_{k}+\mathbf A_k(\mathbf M^{\mathrm s}_{k+1}-\mathbf M^{\mathrm f}_{k+1|k})\mathbf A_k^T$
			\EndFor
		\end{algorithmic}
	\end{algorithm}
	
	\subsection{FG/SPA-Based Algorithm}
	\label{sec:fg_spa}

	\begin{figure*}[!t]
		\centering
		\includegraphics{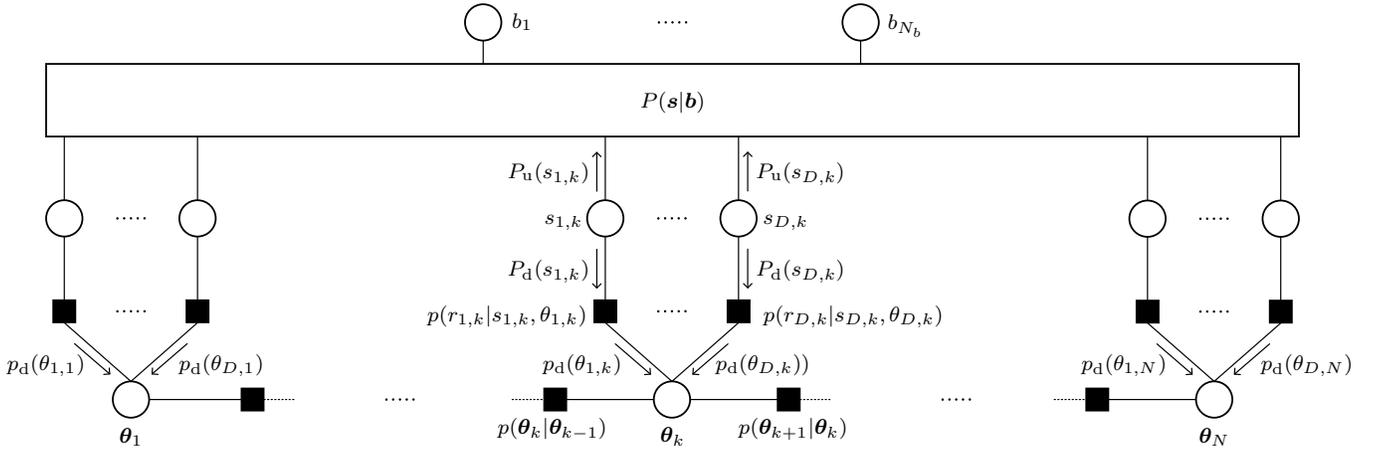}
		\caption{A part of the FG corresponding to \eqref{eq:fg}.}
		\label{fig:fg}
	\end{figure*}
	
	The FG and SPA framework carries out the marginalization of joint distributions in an efficient manner by exploiting how they can be factorized into simpler functions. For a thorough introduction to this framework, refer to \cite{910572}. Moreover, in \cite{1504908}, the framework was applied to the problem of data detection for single-channel satellite transmission in the presence of phase noise. In this paper, we use the framework in a similar manner as in \cite{1504908}, but note that the phase-noise estimation is carried out using a different technique.
	
	The joint distribution $p(\bm b,\bm s,\bm\theta|\bm r)$ in \eqref{eq:marg} factorizes as
	\begin{align}
		&p(\bm b,\bm s,\bm\theta|\bm r)\propto p(\bm b,\bm s,\bm\theta)p(\bm r|\bm b,\bm s,\bm\theta)\nonumber\\
		&= P(\bm s|\bm b)P(\bm b)p(\bm\theta)p(\bm r|\bm s,\bm\theta)\nonumber\\
		&= P(\bm s|\bm b)P(\bm b)p(\bm\theta_1)\prod_{k=2}^N p(\bm\theta_k|\bm\theta_{k-1})\prod_{i,k}p(r_{i,k}|s_{i,k},\theta_{i,k})\nonumber\\
		&\propto P(\bm s|\bm b)\prod_{k=2}^N p(\bm\theta_k|\bm\theta_{k-1})\prod_{i,k}p(r_{i,k}|s_{i,k},\theta_{i,k}),\label{eq:fg}
	\end{align}
	for $i=1,\dots,D$ and $k=1,\dots,N$, where $\propto$ denotes proportionality with respect to $\bm b$, $\bm s$, and $\bm\theta$. Moreover, \eqref{eq:fg} is obtained using the fact that $\bm\theta$ is a random walk, $P(\bm b)$ and $p(\bm\theta_1)$ are uniform distributions, and the received samples are independent of each other given $\bm s$ and $\bm\theta$. Finally, $P(\bm s|\bm b)$ represents the code and mapping constraints, i.e., $P(\bm s|\bm b)$ equals 1 if $\bm s=f(\bm b)$ and 0 otherwise, $p(r_{i,k}|s_{i,k},\theta_{i,k})=\cgaussian{r_{i,k}}{s_{i,k}e^{j\theta_{i,k}}}{2\varN{i}}$ is the likelihood function of $s_{i,k}$ and $\theta_{i,k}$, and $p(\bm\theta_k|\bm\theta_{k-1})=\gaussian{\bm\theta_k}{\bm\theta_{k-1}}{\mathbf Q}$.
	
	The FG associated with \eqref{eq:fg} is shown partially in Fig.~\ref{fig:fg}. In the case of coded transmission where an iterative decoder is utilized, e.g., for low-density parity-check (LDPC) codes and turbo codes, applying the SPA to this FG will yield an iterative phase-noise estimation and decoding algorithm, where in each iteration, the message $\Pd{i,k}{s_{i,k}}$ corresponds to the extrinsic information about $s_{i,k}$ given by the decoder. Moreover, $\Pu{i,k}{s_{i,k}}$ can be regarded as the likelihood function of $s_{i,k}$ from a virtual memoryless phase-noise compensated channel, and is fed to the decoder as bit-wise log-likelihood ratios (LLRs) \cite[Ch.~5.4]{ryan2009channel}. The decoder then either outputs the detected information bits or extrinsic coded-bit LLRs that are converted to $\Pd{i,k}{s_{i,k}}$\footnote{For uncoded transmission, $\Pd{i,k}{s_{i,k}}$ is simply the \textit{a priori} PMF of $s_{i,k}$, and since the FG corresponding to \eqref{eq:fg} does not contain any cycles in the absence of a code \cite{7080901}, the algorithm that results from applying the SPA will not be iterative. As the information bits are assumed equiprobable, $\Pu{i,k}{s_{i,k}}$ is then used to detect the symbols, followed by symbol-to-bit demapping.}. By applying the SPA, it can be shown that
	\begin{align}
		\Pu{i,k}{s_{i,k}}&\propto\int_{\mathbb R^D} p(r_{i,k}|s_{i,k},\theta_{i,k})\frac{p_\mathrm{APP}(\bm\theta_k)}{\pd{i,k}{\theta_{i,k}}}\mathrm d\bm\theta_k,\label{eq:pu}
	\end{align}
	where
	\begin{align}
		p_\mathrm{APP}(\bm\theta_k)&\propto\int_{\mathbb R^{D\times(N-1)}}\prod_{k=2}^N p(\bm\theta_k|\bm\theta_{k-1})\prod_{i,k} \pd{i,k}{\theta_{i,k}}d\bm{\bar\theta}_k
	\end{align}
	represents the \textit{a posteriori} PDF of $\bm\theta_k$, $p(\bm\theta_k|\bm r)$, and $\bm{\bar\theta}_k$ contains all elements of $\bm\theta$ except $\bm\theta_k$. 
	
	\subsubsection{Interface Towards Phase-Noise Estimation}
	The message $p_\mathrm{APP}(\bm\theta_k)$ is approximated using Algorithm \ref{alg:ekf_rtss}. However, the utilization of the EKF requires the likelihood $p(r_{i,k}|\theta_{i,k})$, represented by $\pd{i,k}{\theta_{i,k}}$, to be a complex Gaussian PDF in $r_{i,k}$. For pilot symbols, this is indeed the case since their values are known. For data symbols, however, $\pd{i,k}{\theta_{i,k}}$ is a mixture of complex Gaussian PDFs since
	\begin{align}
		\pd{i,k}{\theta_{i,k}}&=\sum_{s_{i,k}\in\modset}\Pd{i,k}{s_{i,k}}p(r_{i,k}|s_{i,k},\theta_{i,k})\\
		&=\sum_{s_{i,k}\in\modset}\Pd{i,k}{s_{i,k}}\cgaussian{r_{i,k}}{s_{i,k}e^{j\theta_{i,k}}}{2\varN{i}}.
	\end{align}
	To solve this, the same approach is taken as in \cite{1504908}, i.e., $\pd{i,k}{\theta_{i,k}}$ is approximated as a Gaussian PDF with mean and variance chosen such that the Kullback--Leibler (KL) divergence\cite[Ch.~8.5]{Cover2006} between the Gaussian mixture and the single Gaussian is minimized. This yields \cite{Minka:2001}
	\begin{equation}
		\pd{i,k}{\theta_{i,k}}\approx\cgaussian{r_{i,k}}{\bar s_{i,k}e^{j\theta_{i,k}}}{2\varNtilde{i}},\label{eq:pd_approx}
	\end{equation}
	where
	\begin{align}
		\bar s_{i,k}&\define\EVe{P_\mathrm{d}}{S_{i,k}}=\sum_{s_{i,k}\in\modset} s_{i,k}\Pd{i,k}{s_{i,k}},\label{eq:smean}\\
		\varNtilde{i}&\define\varN{i}+\frac12\var{S_{i,k}}\nonumber\\
		&=\varN{i}+\frac12\sum_{s_{i,k}\in\modset} |s_{i,k}-\bar s_{i,k}|^2\Pd{i,k}{s_{i,k}}\label{eq:svar},
	\end{align}
	are soft inputs used by the EKF. For pilot symbols, $\bar s_{i,k}=s_{i,k}$ and $\varNtilde{i}=\varN{i}$, whereas for data symbols, $\bar s_{i,k}$ and $\varNtilde{i}$ are initialized as $\bar s_{i,k}=0$ and $\varNtilde{i}=\varN{i}+E_{\mathrm s}/2$. Due to \eqref{eq:pd_approx}, $\mathbf V_k$ in \eqref{eq:ekf_varcorr} and each component of $\bm h_k$ in \eqref{eq:ekf_corr} are now computed as
	\begin{align}
		\mathbf V_k&=\diag\left(\frac{|\bar s_{1,k}|^2}{\varNtilde{1}},\dots,\frac{|\bar s_{D,k}|^2}{\varNtilde{D}}\right),\label{eq:S_FG}\\
		h_{i,k}&=\frac{1}{\varNtilde{i}}\Im\left\{r_{i,k}\bar s_{i,k}^*e^{-j\hat\theta^{\mathrm f}_{i,k-1}}\right\},\label{eq:h_FG}
	\end{align}
	and the EKF equations are initialized with
	\begin{align}
		\bm{\hat\theta}^{\mathrm f}_{1}&=\left[\angle(r_{1,1}\bar s_{1,1}^*),\dots,\angle(r_{D,1}\bar s_{D,1}^*)\right]^T,\\
		\mathbf M^{\mathrm f}_{1}&=\diag\left(\frac{\varNtilde{1}}{\Es},\dots,\frac{\varNtilde{D}}{\Es}\right).\label{eq:Minit_FG}
	\end{align}
	Hence,
	\begin{equation}
		p_\mathrm{APP}(\bm\theta_k)\approx\gaussian{\theta_{i,k}}{\hat\theta^{\mathrm s}_{i,k}}{M^{\mathrm s}_{i,k}}.\label{eq:pAPPgauss}
	\end{equation}
	
	\subsubsection{Interface Towards Decoder}
	Using \eqref{eq:pAPPgauss}, it is shown in Appendix \ref{app:fgmsg} that $\Pu{i,k}{s_{i,k}}$ can be described approximately in closed form as
	\begin{align}
		\Pu{i,k}{s_{i,k}}&\propto\frac{1}{\sqrt{|\xi_{i,k}(s_{i,k})|}}\exp\left(|\xi_{i,k}(s_{i,k})|-\frac{|s_{i,k}|^2}{2\varN{i}}\right),\label{eq:approxPu}
	\end{align}
	with
	\begin{equation}
		\xi_{i,k}(s_{i,k})\define\frac{e^{j\hat\theta^{\mathrm s}_{i,k}}}{M^{\mathrm s}_{i,k}}+\frac{r_{i,k}s_{i,k}^*}{\varN{i}}-\frac{r_{i,k}\bar s_{i,k}^*}{\varNtilde{i}},\label{eq:xi}
	\end{equation}
	where $M^{\mathrm s}_{i,k}$ denotes the $i$th element on the diagonal line of $\mathbf M^{\mathrm s}_k$. Note that the expression in (\ref{eq:approxPu}) only describes $\Pu{i,k}{s_{i,k}}$ up to a constant and needs to be normalized. Furthermore, it is numerically unstable, and thus, the logarithm of (\ref{eq:approxPu}) can be computed instead, yielding
	\begin{align}
		f_{i,k}(s_{i,k})&\define\ln \Pu{i,k}{s_{i,k}}\nonumber\\
		&\approx|\xi_{i,k}(s_{i,k})|-\frac{\left|s_{i,k}\right|^2}{2\varN{i}}-\frac12\ln|\xi_{i,k}(s_{i,k})|.
	\end{align}
	Finally, $\Pu{i,k}{s_{i,k}}$ is computed from $f_{i,k}(s_{i,k})$ as
	\begin{align}
		\Pu{i,k}{s_{i,k}}=\frac{\exp\left(f_{i,k}(s_{i,k})-f_{i,k}^{\mathrm{max}}\right)}{\sum_{s_{i,k}'\in\modset}\exp\left(f_{i,k}(s_{i,k}')-f_{i,k}^{\mathrm{max}}\right)},\label{eq:f_to_Pu}
	\end{align}
	where $f_{i,k}^{\mathrm{max}}\define\max_{s_{i,k}\in\modset}f_{i,k}(s_{i,k})$. This algorithm will from now on be referred to as FG/SPA-based phase-noise compensation (FG-PNC) and is summarized in Algorithm~\ref{alg:fgspa} for a single iteration.
	
	\begin{algorithm}
		\caption{FG-PNC (1 iteration)}
		\label{alg:fgspa}
		\begin{algorithmic}[1]
			\Require $\bm r,D,N,\mathbf Q,\bm\sigma^2,\modset,\Pd{i,k}{s_{i,k}}~\forall~i,k$
			\Ensure $\Pu{i,k}{s_{i,k}}~\forall~i,k$
			\ForAll{$i,k$}
			\State $\bar s_{i,k}=\sum_{s_{i,k}\in\modset}s_{i,k}\Pd{i,k}{s_{i,k}}$
			\State $\varNtilde{i}=\varN{i}+\frac12\sum_{s_{i,k}\in\modset} |s_{i,k}-\bar s_{i,k}|^2 \Pd{i,k}{s_{i,k}}$
			\EndFor
			\State Compute $(\bm\theta_k^{\mathrm s},\mathbf M_k^{\mathrm s})~\forall~k$ with Alg.~\ref{alg:ekf_rtss} using \eqref{eq:S_FG}--\eqref{eq:Minit_FG}
			\ForAll{$i,k$}
			\ForAll{$s_{i,k}\in\modset$}
			\State $\xi_{i,k}(s_{i,k})=\frac{e^{j\hat\theta_{i,k}^{\mathrm s}}}{M^{\mathrm s}_{i,k}}+\frac{r_{i,k}s_{i,k}^*}{\varN{i}}-\frac{r_{i,k}\bar s_{i,k}^*}{\varNtilde{i}}$
			\State $f_{i,k}(s_{i,k})=|\xi_{i,k}(s_{i,k})|-\frac{\left|s_{i,k}\right|^2}{2\varN{i}}-\frac12\ln|\xi_{i,k}(s_{i,k})|$
			\EndFor
			\State $\Pu{i,k}{s_{i,k}}=\exp(f_{i,k}(s_{i,k})-\max_{s_{i,k}'\in\modset}f_{i,k}(s_{i,k}'))$
			\State Normalize $\Pu{i,k}{s_{i,k}}$ s.t. $\sum_{s_{i,k}\in\modset}\Pu{i,k}{s_{i,k}}=1$
			\EndFor
		\end{algorithmic}
	\end{algorithm}
	
	\subsection{VB-Based Algorithm}
	\label{sec:vb}
	The VB inference framework has been used to develop effective algorithms that perform iterative phase-noise estimation and data detection for single-channel transmission\cite{4799047} and wireless MIMO transmission \cite{7080901}. Here, this framework is exploited in similar way to efficiently solve the marginalization problem in \eqref{eq:marg}. It is worth noting, however, that in \cite[Sec.~IV-C]{4799047}, the VB algorithm is defined for a linearized system model, whereas in this work, we define the algorithm for the system model in \eqref{eq:sysmodel}. To that end, $p(\bm b,\bm s,\bm\theta|\bm r)$ in \eqref{eq:marg} is approximated with a family of factorized distributions, $q_{\bm B,\bm S}(\bm b,\bm s)q_{\bm\Theta}(\bm\theta)$, where $q_{\bm B,\bm S}(\bm b,\bm s)$ and $q_{\bm\Theta}(\bm\theta)$ represent $P(\bm b,\bm s|\bm r)$ and $p(\bm\theta|\bm r)$, resp. The objective is then to minimize the KL divergence between $q_{\bm B,\bm S}(\bm b,\bm s)q_{\bm\Theta}(\bm\theta)$ and $p(\bm b,\bm s,\bm\theta|\bm r)$, i.e.,
	\begin{equation}
		\min_{q_{\bm B,\bm S}(\bm b,\bm s),q_{\bm\Theta}(\bm\theta)}D(q_{\bm B,\bm S}(\bm b,\bm s)q_{\bm\Theta}(\bm\theta)\,||\,p(\bm b,\bm s,\bm\theta|\bm r)).\label{eq:KL}
	\end{equation}
	This minimization is carried out by iteratively updating $q_{\bm B,\bm S}(\bm b,\bm s)$ and $q_{\bm\Theta}(\bm\theta)$, and it can be shown that the update equations at the $l$th iteration are expressed as\cite[Ch.~2.3]{beal:varbayes}
	\begin{align}
		\qGamma{\bm\Theta}{l}{\bm\theta}&\propto\exp\left(\EVe{\qSi{\bm B,\bm S}{l-1}}{\log p(\bm\theta,\bm r|\bm B,\bm S)}\right)\nonumber\\
		&=p(\bm\theta)\exp\left(\EVe{\qSi{\bm S}{l-1}}{\log p(\bm r|\bm S,\bm\theta)}\right)\label{eq:q_gamma},\\
		\qS{\bm B,\bm S}{l}{\bm b,\bm s}&\propto P(\bm b,\bm s)\exp\left(\EVe{\qGammai{\bm\Theta}{l}}{\log p(\bm\Theta,\bm r|\bm b,\bm s)}\right)\nonumber\\
		&\propto P(\bm s|\bm b)\exp\left(\EVe{\qGammai{\bm\Theta}{l}}{\log p(\bm r|\bm s,\bm\Theta)}\right)\label{eq:q_s},
	\end{align}
	for $l=1,\dots,\Nit$, where $\Nit$ is the number of iterations and $p(\bm r|\bm s,\bm\theta)$ is the likelihood function of $\bm s$ and $\bm\theta$. These updates give rise to an algorithm that performs iterative decoding and phase-noise estimation. For more details on this framework\footnote{VB inference can also be regarded as a specific type of message passing on FGs, often referred to as variational message passing in that context. See \cite{4557602} for more details and \cite{5683839} for an example that uses this concept for joint channel estimation and decoding in wireless MIMO transmission.} and on the results in \eqref{eq:q_gamma} and \eqref{eq:q_s}, refer to \cite{4799047,beal:varbayes}.
	The rest of the subsection details the computation of $\qGamma{\bm\Theta}{l}{\bm\theta}$ and $\qS{\bm B,\bm S}{l}{\bm b,\bm s}$ for one iteration, and therefore, the superscript indicating the iteration number will be skipped hereafter for notational convenience.
	
	Since all received samples are statistically independent of each other given $\bm s$ and $\bm\theta$, \eqref{eq:q_gamma} and \eqref{eq:q_s} can be rewritten as
	\begin{align}
		\qGammaSI{\bm\Theta}{\bm\theta}\propto{}&p(\bm\theta)\nonumber\\
		&\cdot\prod_{i,k}\exp\left(-\frac{1}{2\sigma_i^2}\EVe{\qSiSI{S_{i,k}}}{\left|r_{i,k}-S_{i,k}e^{j\theta_{i,k}}\right|^2}\right),\label{eq:qGammaExp}\\
		\qSSI{\bm B,\bm S}{\bm b,\bm s}\propto{}&P(\bm s|\bm b)\nonumber\\
		&\cdot\prod_{i,k}\exp\left(-\frac{1}{2\sigma_i^2}\EVe{\qGammaiSI{\Theta_{i,k}}}{\left|r_{i,k}-s_{i,k}e^{j\Theta_{i,k}}\right|^2}\right),\label{eq:qSexp}
	\end{align}
	for $i=1,\dots,D$ and $k=1,\dots,N$. It can be seen from \eqref{eq:qGammaExp} that $\qGammaSI{\bm\Theta}{\bm\theta}$ relies on $\qSSI{S_{i,k}}{s_{i,k}}$, the marginals of $\qSSI{\bm B,\bm S}{\bm b,\bm s}$. Likewise, \eqref{eq:qSexp} shows that $\qSSI{\bm S}{\bm s}$ relies on $\qGammaSI{\Theta_{i,k}}{\theta_{i,k}}$, the marginals of $\qGammaSI{\bm\Theta}{\bm\theta}$. Thus, computing \eqref{eq:q_gamma} and \eqref{eq:q_s} involves the marginalization of $\qGammaSI{\bm\Theta}{\bm\theta}$ and $\qSSI{\bm S}{\bm s}$.
	
	\subsubsection{Interface Towards Phase-Noise Estimation}
	The expression in \eqref{eq:qGammaExp} can be simplified by noting that
	\begin{align}
		&\EVe{\qSiSI{S_{i,k}}}{\left|r_{i,k}-S_{i,k}e^{j\theta_{i,k}}\right|^2}\nonumber\\
		&=\left|r_{i,k}-\EVe{\qSiSI{S_{i,k}}}{S_{i,k}}e^{j\theta_{i,k}}\right|^2+\var{S_{i,k}}\label{eq:qGammaVar}\\
		&=\left|r_{i,k}-\bar{\bar s}_{i,k}e^{j\theta_{i,k}}\right|^2+\var{S_{i,k}},
	\end{align}
	where
	\begin{equation}
		\bar{\bar s}_{i,k}\define\EVe{\qSiSI{S_{i,k}}}{S_{i,k}}=\sum_{s_{i,k}\in\modset}s_{i,k}\qSSI{S_{i,k}}{s_{i,k}},\label{eq:smeanVB}
	\end{equation}
	and \eqref{eq:qGammaVar} uses the fact that $\EVi{|\cdot|^2}=\var{\cdot}+|\EVi{\cdot}|^2$. Similarly to \eqref{eq:smean}, $\bar{\bar s}_{i,k}$ represents a soft input that is utilized in the phase-noise estimation. Thus,
	\begin{align}
		\qGammaSI{\bm\Theta}{\bm\theta}\propto{}&p(\bm\theta)\nonumber\\
		&\cdot\prod_{i,k}\exp\left(-\frac{1}{2\sigma_i^2}\left(\left|r_{i,k}-\bar{\bar s}_{i,k}e^{j\theta_{i,k}}\right|^2+\var{S_{i,k}}\right)\right)\nonumber\\
		\propto{}& p(\bm\theta)\prod_{i,k}\exp\left(-\frac{1}{2\sigma_i^2}\left|r_{i,k}-\bar{\bar s}_{i,k}e^{j\theta_{i,k}}\right|^2\right)\nonumber\\
		\propto{}&\prod_{k=2}^N p(\bm\theta_k|\bm\theta_{k-1})\prod_{i,k}\cgaussian{r_{i,k}}{\bar{\bar s}_{i,k}e^{j\theta_{i,k}}}{2\varN{i}}.\label{eq:qGamma}
	\end{align}
	Analogously to FG-PNC, $\bar{\bar s}_{i,k}=s_{i,k}$ for pilot symbols during each iteration, and for data symbols, $\bar{\bar s}_{i,k}$ is initialized as 0. The structure of  $\qGammaSI{\bm\Theta}{\bm\theta}$ in \eqref{eq:qGamma} allows for its marginals, $\qGammaSI{\bm\Theta_k}{\bm\theta_k}$, to be approximated using Algorithm~\ref{alg:ekf_rtss}. The means and covariances of these marginals are computed with \eqref{eq:ekf_varpred}--\eqref{eq:ekf_corr} and \eqref{eq:gammaA}--\eqref{eq:gammahatvar}. Due to \eqref{eq:qGamma}, $\mathbf V_k$ in \eqref{eq:ekf_varcorr} and the components of $\bm h_k$ in \eqref{eq:ekf_corr} are computed as
	\begin{align}
		\mathbf V_k&=\diag\left(\frac{|\bar{\bar s}_{1,k}|^2}{\varN{1}},\dots,\frac{|\bar{\bar s}_{D,k}|^2}{\varN{D}}\right),\label{eq:S_VB}\\
		h_{i,k}&=\frac{1}{\varN{i}}\Im\left\{r_{i,k}\bar{\bar s}_{i,k}^*e^{-j\hat\theta^{\mathrm f}_{i,k-1}}\right\},\label{eq:h_VB}
	\end{align}
	and the EKF equations are initialized with
	\begin{align}
		\bm{\hat\theta}^{\mathrm f}_{1}&=\left[\angle(r_{1,1}\bar{\bar s}_{1,1}^*),\dots,\angle(r_{D,1}\bar{\bar s}_{D,1}^*)\right]^T,\\
		\mathbf M^{\mathrm f}_{1}&=\diag\left(\frac{\varN{1}}{\Es},\dots,\frac{\varN{D}}{\Es}\right).\label{eq:Minit_VB}
	\end{align}
	Further marginalizing $\qGammaSI{\bm\Theta_k}{\bm\theta_k}$ to obtain $\qGammaSI{\Theta_{i,k}}{\theta_{i,k}}$ is trivial since $\qGammaSI{\bm\Theta_k}{\bm\theta_k}$ is approximated as a Gaussian PDF \cite[Ch.~8]{Rasmussen:2006:GPM:1162254}, yielding
	\begin{equation}
		\qGammaSI{\Theta_{i,k}}{\theta_{i,k}}\approx\gaussian{\theta_{i,k}}{\hat\theta^{\mathrm s}_{i,k}}{M^{\mathrm s}_{i,k}}.\label{eq:qGammaGauss}
	\end{equation}
	
	\subsubsection{Interface Towards Decoder}
	The expectation in \eqref{eq:qSexp} can be expressed as
	\begin{align}
		&\EVe{\qGammaiSI{\Theta_{i,k}}}{\left|r_{i,k}-s_{i,k}e^{j\Theta_{i,k}}\right|^2}\nonumber\\
		&=|r_{i,k}|^2-2\Re\left\{r_{i,k}s_{i,k}^*\alpha_{i,k}^*\right\}+|s_{i,k}|^2,\label{eq:EVqGamma}
	\end{align}
	where $\alpha_{i,k}\define\EVe{\qGammaiSI{\Theta_{i,k}}}{e^{j\Theta_{i,k}}}$. To compute $\alpha_{i,k}$ in closed form, note that the characteristic function of a real Gaussian random variable $X$ with mean $\mu$ and variance $\sigma^2$ is $\Phi_X(\omega)\define\EVe{X}{e^{j\omega X}}=\exp(j\mu\omega-\omega^2\sigma^2/2)$ \cite[Ch.~4.7]{miller:probability}. Using this fact as well as \eqref{eq:qGammaGauss}, $\alpha_{i,k}$ can be described as $\Phi_{\Theta_{i,k}}(1)=\exp(j\hat\theta^{\mathrm s}_{i,k}-M^{\mathrm s}_{i,k}/2)$. Moreover, \eqref{eq:EVqGamma} leads to
	\begin{align}
		\qSSI{\bm B,\bm S}{\bm b,\bm s}&\propto P(\bm s|\bm b)\prod_{i,k}\exp\left(\frac{\Re\{r_{i,k}s_{i,k}^*\alpha_{i,k}^*\}}{\sigma_i^2}-\frac{|s_{i,k}|^2}{2\sigma_i^2}\right)\nonumber\\
		&\propto P(\bm s|\bm b)\prod_{i,k} g_{i,k}(s_{i,k}),\label{eq:qS_g}
	\end{align}
	where $g_{i,k}(s_{i,k})$ corresponds to the likelihood function of $s_{i,k}$ from a virtual memoryless phase-noise compensated channel, analogously to $\Pu{i,k}{s_{i,k}}$ in \eqref{eq:approxPu}. To obtain $\qSSI{S_{i,k}}{s_{i,k}}$, which is used to compute $\qGammaSI{\bm\Theta}{\bm\theta}$ in \eqref{eq:qGammaExp} during a consecutive iteration, $\qSSI{\bm B,\bm S}{\bm b,\bm s}$ is marginalized over all bits and all symbols except $s_{i,k}$. The marginalization of $\qSSI{\bm B,\bm S}{\bm b,\bm s}$ is performed in the decoder, where the function $g_{i,k}(s_{i,k})$ in \eqref{eq:qS_g} is converted to bit-wise LLRs that are fed to the decoder. The decoder then either outputs the detected information bits or \textit{a posteriori} coded-bit LLRs that are converted to $\qSSI{S_{i,k}}{s_{i,k}}$.\footnote{For uncoded transmission, this marginalization is trivial since the transmitted symbols are independent of each other. Thus, $\qSSI{S_{i,k}}{s_{i,k}}\propto P(s_{i,k})g_{i,k}(s_{i,k})$, where $P(s_{i,k})$ is a uniform PMF for data symbols and a degenerate distribution for pilot symbols, i.e., equal to 1 if $s_{i,k}$ is equal to the pilot and 0 otherwise.}
	This algorithm will be referred to as VB-based phase-noise compensation (VB-PNC) and is summarized in Algorithm~\ref{alg:vb}.
	\begin{algorithm}
		\caption{VB-PNC (1 iteration)}
		\label{alg:vb}
		\begin{algorithmic}[1]
			\Require $\bm r, D, N, \mathbf Q, \bm\sigma^2, \modset, \qS{S_{i,k}}{l-1}{s_{i,k}}~\forall~i,k$
			\Ensure $g_{i,k}^{(l)}(s_{i,k})~\forall~i,k$
			\ForAll{$i,k$}
			\State $\bar{\bar s}_{i,k}=\sum_{s_{i,k}\in\modset}s_{i,k}\qS{S_{i,k}}{l-1}{s_{i,k}}$
			\EndFor
			\State Compute $(\bm\theta_k^{\mathrm s},\mathbf M_k^{\mathrm s})~\forall~k$ with Alg.~\ref{alg:ekf_rtss} using \eqref{eq:S_VB}--\eqref{eq:Minit_VB}
			\ForAll{$i,k$}
			\State $\alpha_{i,k}=\exp(j\hat\theta^{\mathrm s}_{i,k}-{M^{\mathrm s}_{i,k}}/{2})$
			\ForAll{$s_{i,k}\in\modset$}
			\State 	$g_{i,k}(s_{i,k})=\exp\left(\frac{\Re\left\{r_{i,k}s_{i,k}^*\alpha_{i,k}^*\right\}}{\sigma_i^2}-\frac{|s_{i,k}|^2}{2\sigma_i^2}\right)$
			\EndFor
			\State Normalize $g_{i,k}(s_{i,k})$ s.t. $\sum_{s_{i,k}\in\modset}g_{i,k}(s_{i,k})=1$
			\EndFor
		\end{algorithmic}
	\end{algorithm}
	
	\textit{Remark 1:} It can be seen from \eqref{eq:smean} and \eqref{eq:smeanVB} that in the absence of pilot symbols, FG-PNC and VB-PNC will not bootstrap properly. However, it is shown in \cite{7130553} that trellis-based demodulation can be used to bootstrap the Tikhonov algorithm in \cite[Sec.~IV-B]{1504908} without pilot symbols. We conjecture that the same principles could be used for FG-PNC and VB-PNC; however, such a study is out of scope for this paper and is suggested for future work.
	
	\textit{Remark 2:} FG-PNC takes into account the symbol uncertainty when interfacing with the phase-noise estimation, in contrast to VB-PNC. This can be seen from the approximate likelihoods in \eqref{eq:pd_approx} and \eqref{eq:qGamma}. However, both algorithms convey the phase-noise uncertainty to the decoder, which can be seen from \eqref{eq:approxPu} and \eqref{eq:qS_g}. Moreover, VB-PNC can perform iterative phase-noise estimation and data detection even in the case of uncoded transmission. As mentioned in Section~\ref{sec:fg_spa}, the FG/SPA framework does not yield an iterative algorithm in the absence of an iterative decoder.	However, multiple iterations can be performed with FG-PNC for uncoded transmission by using $\Pu{i,k}{s_{i,k}}$ instead of $\Pd{i,k}{s_{i,k}}$ to compute \eqref{eq:smean} and \eqref{eq:svar} in a consecutive iteration.
	
	\subsection{Distribution of Pilot Symbols}
	The optimal distribution of the pilot symbols for multichannel transmission, subject to an average pilot rate across the channels, in nontrivial in general. For single-channel transmission, equispaced pilot symbols typically give good performance. However, an effective pilot distribution for multichannel transmission cannot be easily inferred from studying the single-channel case, as the optimal distribution will depend on the phase-noise correlation across channels, as well as the SNR. In \cite{alfredsson2017ecoc}, we investigated this problem for SDM transmission and found that arranging the pilot symbols on a wrapped diagonal in the space--time grid performs well in general. This type of distribution will be used for the remainder of this paper.
	
	\subsection{Conversion Between PMFs and LLRs}
	As detailed in the previous subsections, the inputs and outputs to FG-PNC and VB-PNC in each iteration are in the form of symbol PMFs. However, iterative decoders for binary LDPC codes and turbo codes are typically implemented in the logarithm domain \cite[Ch.~5]{ryan2009channel}, and thus, have bit-wise LLRs as inputs and outputs.
	
	The computation of bit-wise input LLRs is done as follows. For each transmitted symbol $s_{i,k}$, denote with $c_{i,k}^j$ the $j$th coded bit in the binary labeling of the constellation points. The LLR for $c_{i,k}^j$ is computed from the likelihood of $s_{i,k}$ as
	\begin{align}
		L(c_i)&\define\log\left(\frac{p(r_{i,k}|c_{i,k}^j=0)}{p(r_{i,k}|c_{i,k}^j=1)}\right)\nonumber\\
		&=\log\left(\frac{\sum_{s_{i,k}\in\mathcal B_0}p(r_{i,k}|s_{i,k})}{\sum_{s_{i,k}\in\mathcal B_1}p(r_{i,k}|s_{i,k})}\right),
	\end{align}
	for $j=1,\dots,R_\mathrm{m}$, where $R_\mathrm{m}\define\log_2|\modset|$, and $\mathcal B_\nu$ is the set of constellation points that have the $j$th bit in the binary labeling as $\nu\in\{0,1\}$. In relation to the derived algorithms, $p(r_{i,k}|s_{i,k})$ corresponds to $\Pu{i,k}{s_{i,k}}$ for FG-PNC and $g_{i,k}(s_{i,k})$ for VB-PNC.
	
	The \textit{a posteriori} output LLRs from the decoder are defined as
	\begin{align}
		L(c_{i,k}^j|\bm r)&\define\log\left(\frac{P(c_{i,k}^j=0|\bm r)}{P(c_{i,k}^j=1|\bm r)}\right),\label{eq:outLLR}
	\end{align}
	To compute $\qSSI{S_{i,k}}{s_{i,k}}$ for VB-PNC, the output LLRs in \eqref{eq:outLLR} are first converted to \textit{a posteriori} coded-bit probabilities as
	\begin{align}
		P(c_{i,k}^j=0|\bm r)&=\frac{e^{L(c_{i,k}^j|\bm r)}}{1+e^{L(c_{i,k}^j|\bm r)}}.\label{eq:outP}
	\end{align}
	Then, with a slight abuse of notation, denote the probability of $s_{i,k}$ being a constellation point with binary labeling $(\nu_1,\dots,\nu_{R_\mathrm{m}})\in\{0,1\}^{R_\mathrm{m}}$ as $P(s_{i,k}=(\nu_1,\dots,\nu_{R_\mathrm{m}}))$, which corresponds to $\qSSI{S_{i,k}}{s_{i,k}}$ and is computed as
	\begin{equation}
		P(s_{i,k}=(\nu_1,\dots,\nu_{R_\mathrm{m}}))=\prod_{j=1}^{R_\mathrm{m}}P(c_{i,k}^j=\nu_j|\bm r).\label{eq:sOutP}
	\end{equation}
	For FG-PNC, $\Pd{i,k}{s_{i,k}}$ is found in the same way as $\qSSI{S_{i,k}}{s_{i,k}}$, except the coded-bit probabilities in \eqref{eq:sOutP} are computed from the extrinsic LLRs $L_\mathrm{e}(c_{i,k}^j)\define L(c_{i,k}^j|\bm r)-L(c_{i,k}^j)$.
	
	\subsection{Computational Complexity}
	The main difference in computational complexity (in terms of the number of real additions and multiplications per outer iteration between the phase-noise compensation and decoding) of the two algorithms comes from FG-PNC computing the symbol uncertainties in \eqref{eq:svar}, which VB-PNC does not. This computation scales as $\mathcal O(DN|\modset|)$. Both algorithms use the same method to estimate the marginal phase-noise PDFs, which scales as $\mathcal O(D^3N)$ due to matrix inversions, and the likelihood computations in \eqref{eq:f_to_Pu} and \eqref{eq:qS_g} both scale as $\mathcal O(DN|\modset|)$. It is important to mention that this method of quantifying complexity only gives an approximate view of the actual hardware requirements needed to implement the algorithms. Hence, it serves as a starting point for a more detailed analysis, which is out of scope for this paper.
	
	\section{Performance Results}
	In this section, we first justify the utilization of the EKF by showing that the linearization of the system model in \eqref{eq:sysmodel} does not yield any significant penalties to the phase-noise estimation for practical baud rates and laser linewidths.
	Thereafter, we experimentally validate the system model in \eqref{eq:sysmodel} and the proposed algorithms. We then further assess their performance using Monte Carlo simulations.
	
	\subsection{Impact of EKF Linearization}
	\label{sec:linerror}
\begin{figure}[!t]
		\centering
		\includegraphics{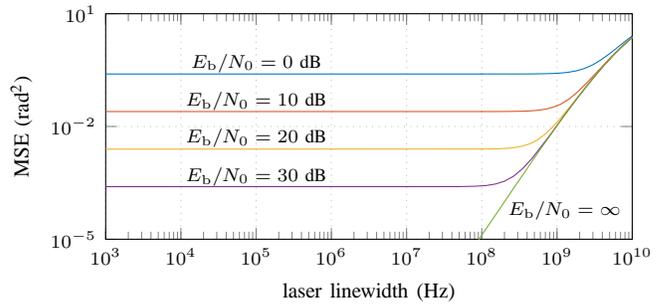}
		\caption{MSE of a phase-noise estimate, obtained using a linearized system model, versus laser linewidth.}
		\label{fig:linError}
	\end{figure}

	The linearization of the system model, which is imposed by the EKF, can negatively impact the phase-noise estimation \cite{6212402,6422412,8253485} if the phase noise varies fast enough. In \cite{6422412}, this effect is investigated for oscillator phase noise in wireless MIMO systems and an error floor is attributed to linearization penalties. However, for all parameters of interest in fiber-optical systems, the linearization has negligible effects on the phase-noise estimation performance. To show this, the mean squared error (MSE) is computed from $10^6$ realizations of a phase-noise estimate that is obtained at time $k$ from a received sample using \eqref{eq:g}, assuming perfect knowledge of the transmitted symbol and the phase noise at time $k-1$. For 20 GBaud uncoded transmission, the resulting MSE is shown in Fig.~\ref{fig:linError} as a function of laser linewidth, for different SNRs per information bit, defined as
	\begin{equation}
		\snr\define\frac{\Es}{2\sigma^2 R_\mathrm{c}R_\mathrm{m}(1-R_\mathrm{p})},
	\end{equation}
	where $\sigma^2$ is the variance per real dimension of the complex AWGN, and $R_\mathrm{p}$, $R_\mathrm{c}$, and $R_\mathrm{m}$ are the pilot rate, code rate, and bits per symbol, resp. From Fig.~2 it can be seen that for laser linewidths under $10^8$ Hz, the MSE in the absence of AWGN (and thus, due to the linearization) is below $10^{-5}$, while in the presence of AWGN, the MSE is orders of magnitude larger, even at high $\snrinline$. Hence, for practical laser linewidths (on the order of MHz and below) and SNRs, the total MSE is virtually only due to the AWGN; in other words, the linearization yields negligible estimation error. The choice of modulation format and pilot rate has a marginal effect on the results.
	
	\subsection{Experimental Verification}
	\label{sec:exp_ver}
	
\begin{figure}[!t]
		\centering
		\includegraphics{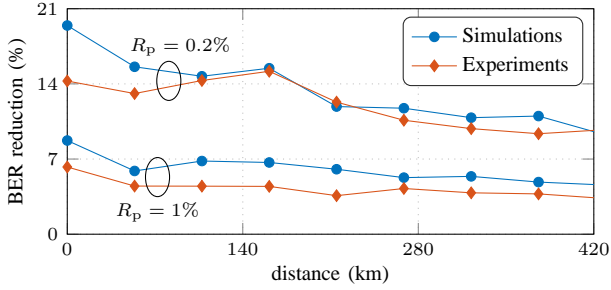}
		\caption{Reduction in BER by performing joint-channel processing instead of per-channel processing for different distances of uncoded 64QAM transmission, comparing the results using experimental data and simulations based on the system model in \eqref{eq:sysmodel} for two pilot rates.}
		\label{fig:expVSsim}
	\end{figure}

	To assess the efficacy of the system model in \eqref{eq:sysmodel} and the proposed algorithms, a recirculating-loop experiment involving SDM transmission of 64QAM at 20 GBaud through 3 cores of a 7-core weakly-coupled, homogeneous, multicore fiber was carried out. The experimental setup and detailed results are given in a separate article \cite{8576586}. In this section, additional results are presented based on the same data set. The DSP chain prior to phase-noise compensation consisted of standard steps, applied on a per-core or per-channel basis. The results in Fig.~\ref{fig:expVSsim} were obtained for two pilot rates using a symbol-detection algorithm analogous to FG-PNC, performing both per-channel and joint-channel processing. As can be seen, jointly processing the channels leads to a reduction in BER relative to what is attained with per-channel processing. The amount of BER reduction is greater for lower transmission distances, and also increases when the pilot rate is decreased.
		
	Per-channel and joint-channel processing is also performed on realizations of \eqref{eq:sysmodel} using the estimated SNR values from the experimental data, as well as a particular configuration of $\mathbf Q$ that was used when processing the experimental data. The results are shown in Fig.~\ref{fig:expVSsim}. A strong agreement is observed overall between the results based on experiments and simulations, which suggests that the system model in \eqref{eq:sysmodel} and the proposed algorithms are relevant in this application. It is, however, important to note that other types of multichannel optical systems could have drastically different behavior, in which case other models and algorithms may be more suitable than what has been investigated in this paper.
	
	\begin{figure*}[!t]
		\centering
		\includegraphics{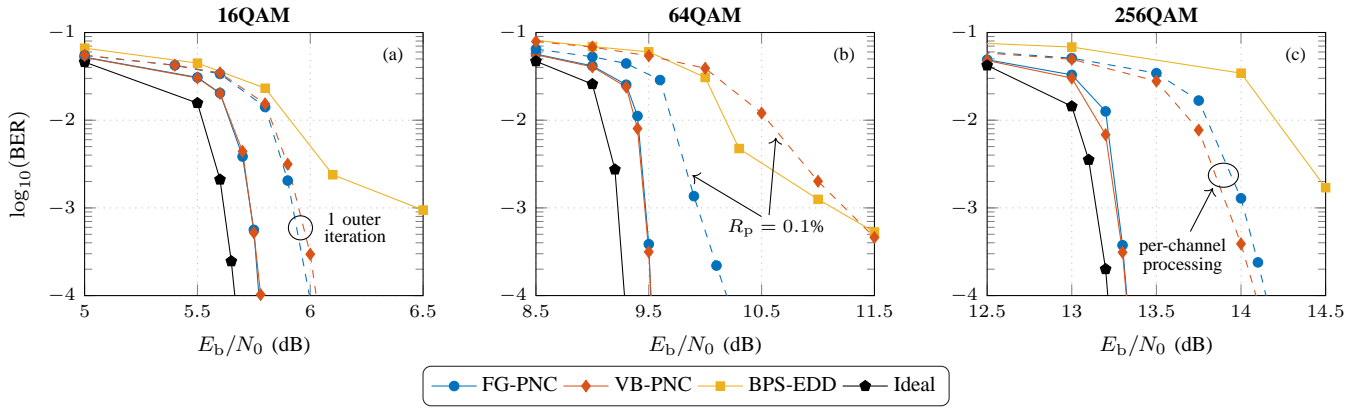}
		\caption{BER versus SNR per information bit for 16QAM and 1 MHz laser linewidth (a), 64QAM and 1 MHz laser linewidth (b), and 256QAM and 100 kHz laser linewidth (c), assuming a symbol rate of 20 GBaud and $R_\mathrm{p}=1\%$.}
		\label{fig:results}
	\end{figure*}
	
	\subsection{Simulation Results}
	\label{sec:simresults}
	The proposed algorithms are further assessed in terms of phase-noise tolerance through Monte Carlo simulations. In order to put the results into perspective, their performance is compared to a strategy entailing per-channel phase-noise compensation using BPS and symbol detection based on minimum Euclidean distance (BPS-EDD). Coded transmission of 16QAM, 64QAM, and 256QAM over 20 channels is considered, using $R_\mathrm{p}=1\%$ for the proposed algorithms. An off-the-shelf rate-4/5 LDPC code from the DVB-S.2 standard is utilized to encode information bits independently for each channel, yielding codewords of length 64800 bits. Furthermore, the variance of the AWGN is kept identical for all channels and the phase noise has a high degree of correlation across the channels. More specifically, the covariance matrix of $\Delta\bm\theta_k$ in \eqref{eq:random_walk}, $\mathbf Q$, is constructed such that the elements on the diagonal are equal to $\varLPN+\varDrift$, with $\varDrift\ll\varLPN$, while all other elements are equal to $\varLPN$, where $\varLPN=2\pi\Delta\nu T_\mathrm{s}$ is the laser phase noise variance and is a function of the laser linewidth, $\Delta\nu$, and the symbol duration, $T_\mathrm{s}$. This corresponds to a single dominant laser phase noise component that is common to all channels, in addition to phase drifts that are independent between channels. As these drifts are typically orders of magnitude slower than the laser phase noise \cite{6317137}, we set $\varDrift=\varLPN/1000$. The laser linewidth and symbol duration product, $\Delta\nu T_\mathrm{s}$, is fixed at $5\cdot10^{-5}$ for 16QAM and 64QAM, and $5\cdot10^{-6}$ for 256QAM. Assuming a 20 GBaud symbol rate, this corresponds to linewidths of 1 MHz and 100kHz, resp. Perfect knowledge of $\mathbf Q$ at the receiver is assumed. BER estimates are obtained for different $\snrinline$ values by counting a minimum of 100 frame errors. Furthermore, the BER performance of coded transmission with ideal phase-noise compensation in the absence of pilots is included as a benchmark.
	
	The scheduling for the proposed algorithms is as follows. A total of 2 outer iterations are performed, where in each outer iteration, the decoder is run for 50 decoding iterations. The information bits are then detected following the second outer iteration. Due to limitations in the decoder implementation, the decoder state is reset between each outer iteration. This incurs a performance penalty for FG-PNC as the message passing that results from applying the SPA to the FG in Fig.~\ref{fig:fg} does not assume resets in the decoder state.
	
	When performing phase-noise compensation using BPS, 32 test phases are used for transmission of 16QAM, whereas 64 test phases are used for 64QAM and 256QAM. A filter length between 70 and 90 is used. No differential encoding is used, but instead, the initial value of the phase noise is assumed to be known. Moreover, only a single outer iteration is run since BPS does not exploit the symbol statistics. Thus, following BPS and symbol detection, the decoder performs 50 decoding iterations and outputs the detected information bits.
	
	Figs.~\ref{fig:results}~(a), (b), and (c) show results for 16QAM, 64QAM, and 256QAM, resp., for FG-PNC, VB-PNC, and BPS-EDD. In all cases, FG-PNC and VB-PNC perform similarly and outperform BPS-EDD by a wide margin. Compared to the ideal performance, they result in a negligible SNR penalty of 0.1 dB, 0.2 dB, and 0.1 dB for 16QAM, 64QAM, and 256QAM, resp., at a BER of $10^{-4}$. In addition, Fig.~\ref{fig:results}~(a) includes results for FG-PNC and VB-PNC running 1 only outer iteration, showing how the cooperation between the decoder and the phase-noise estimation improves BER performance. Analogous results are also obtained for 64QAM and 256QAM; however, for clarity, they are not shown in the plots.
	Fig.~\ref{fig:results}~(b) shows additional results for FG-PNC and VB-PNC using $R_\mathrm{p}=0.1\%$. In this case, FG-PNC demonstrates superior performance to VB-PNC, which is found to be mainly due to the different likelihood approximations in \eqref{eq:approxPu} and \eqref{eq:qS_g}. Similar results are also found for 16QAM and 256QAM.
	Finally, Fig.~\ref{fig:results}~(c) contains results when the proposed algorithms are used for per-channel processing. As expected, this yields worse performance than joint processing of all channels. Again, comparable results are found for 16QAM and 64QAM.
	
	The strong performance of the proposed algorithms can be attributed to the following: (i) Due to the phase-noise correlation, the pilot symbol distribution allows for a more effective use of the pilots in joint-channel processing compared to per-channel processing. In addition, resistance to AWGN is improved through joint-channel processing; (ii) Furthermore, the algorithms make use of the phase-noise statistics when computing the symbol likelihoods, which has been shown to be a superior strategy to separating the phase-noise compensation and symbol detection \cite{310614}; (iii) Finally, the iterative cooperation with the decoder improves the phase-noise compensation performance, and thus, the resulting BER. As expected, however, the amount of performance improvement diminishes with increasing number of outer iterations.
		
	\section{Discussion and Conclusions}
	\label{sec:conclusions}
	
	This work was motivated by the fact that in certain multichannel optical systems, the received signal may be approximated by a simplistic multidimensional phase-noise model after being processed by standard DSP steps prior to phase-noise compensation. The MAP bit detector for this model was approximated using two frameworks, and the resulting pilot-aided algorithms perform iterative phase-noise estimation in cooperation with an iterative decoder. By exploiting the phase-noise correlation across the channels, a more capable phase-noise compensation is achieved than what is possible with per-channel processing. The system model and the proposed algorithms were verified through comparison of results based on experimental data and Monte Carlo simulations. Moreover, the algorithms were further assessed using simulations in terms of phase-noise tolerance for different modulation formats, amounts of phase noise, and pilot rates. The results in Fig.~\ref{fig:results} show that the proposed algorithms significantly outperform the typical phase-noise compensation approach in the optical literature, giving rise to a marginal SNR penalty of 0.2 dB or less at a BER of $10^{-4}$ with respect to pilot-free transmission in the absence of phase noise. However, aside from the increased transmission reach shown in Fig.~\ref{fig:expVSsim}, practical implications have not been addressed in this paper. For example, can joint-channel processing increase power and spectral efficiency, or can hardware requirements be relaxed? These are important questions that we have addressed in \cite{8576586}.
	
	\appendices
	\section{Derivation of EKF Equations}
	\label{app:ekf_derivation}
	The EKF equations in Sec.~\ref{sec:ekf_rtss} can be derived from the general recursive Bayesian filtering equations \cite[Ch.~4.2]{sarkka:bayesian}
	\begin{align}
		&p(\bm\theta_k|\bm r_1,\dots,\bm r_k)\nonumber\\
		&\propto p(\bm r_k|\bm\theta_k)\int_{\mathbb R^D} p(\bm\theta_k|\bm\theta_{k-1})p(\bm\theta_{k-1}|\bm r_1,\dots,\bm r_{k-1})d\bm\theta_{k-1},\label{eq:bayes_rec}
	\end{align}
	for $k=2,\dots,N$. Conforming to the EKF, the system model in \eqref{eq:sysmodel} at time $k$ is linearized using a first-order Taylor expansion around an estimate of $\bm\theta_{k-1}$, denoted with $\bm{\hat\theta}^{\mathrm f}_{k-1}$, yielding
	\begin{equation}
		r_{i,k}\approx s_{i,k}e^{j\hat\theta^{\mathrm f}_{i,k-1}}(1+j(\theta_{i,k}-\hat\theta^{\mathrm f}_{i,k-1}))+n_{i,k},
	\end{equation}
	where $\hat\theta^{\mathrm f}_{i,k-1}$ is the $i$th element of $\bm{\hat\theta}^{\mathrm f}_{k-1}$. Hence,
	\begin{align}
		&p(\bm r_k|\bm\theta_k)\nonumber\\
		&=\prod_{i=1}^D p(r_{i,k}|\theta_{i,k})\nonumber\\
		&=\prod_{i=1}^D \cgaussian{r_{i,k}}{s_{i,k}e^{j\theta_{i,k}}}{2\varN{i}}\label{eq:known_s_likelihood}\\
		&\approx\prod_{i=1}^D\cgaussian{r_{i,k}}{s_{i,k}e^{j\hat\theta^{\mathrm f}_{i,k-1}}(1+j(\theta_{i,k}-\hat\theta^{\mathrm f}_{i,k-1}))}{2\varN{i}}\nonumber\\
		&\propto\prod_{i=1}^D\exp\left(-\frac{|s_{i,k}|^2}{2\varN{i}}\left|j+\hat\theta^{\mathrm f}_{i,k-1}+\eta_{i,k}-\theta_{i,k}\right|^2\right)\label{eq:app_ekf1}\\
		&\propto\prod_{i=1}^D\exp\left(-\frac{|s_{i,k}|^2}{2\varN{i}}\left(\hat\theta^{\mathrm f}_{i,k-1}+\Re\{\eta_{i,k}\}-\theta_{i,k}\right)^2\right)\label{eq:app_ekf2}\\
		&\propto\prod_{i=1}^D\gaussian{\theta_{i,k}}{\hat\theta^{\mathrm f}_{i,k-1}+\Im\left\{\frac{r_{i,k}e^{-j\hat\theta^{\mathrm f}_{i,k-1}}}{s_{i,k}}\right\}}{\frac{\varN{i}}{|s_{i,k}|^2}},\label{eq:app_ekf3}
	\end{align}
	where $\eta_{i,k}\define r_{i,k}e^{-j\hat\theta^{\mathrm f}_{i,k-1}}/(js_{i,k})$. Furthermore, \eqref{eq:known_s_likelihood} follows as the knowledge of $s_{i,k}$ for all $i$ and $k$ is assumed, and \eqref{eq:app_ekf2} is obtained by using $|z|^2=\Re\{z\}^2+\Im\{z\}^2$, for $z\in\mathbb C$, as
	\begin{align}
		&\left|j+\hat\theta^{\mathrm f}_{i,k-1}+\eta_{i,k}-\theta_{i,k}\right|^2\nonumber\\
		&=\left(1+\Im\{\eta_{i,k}\}\right)^2+\left(\hat\theta^{\mathrm f}_{i,k-1}+\Re\{\eta_{i,k}\}-\theta_{i,k}\right)^2.\label{eq:app_ekf4}
	\end{align}
	The first term in \eqref{eq:app_ekf4} is constant with respect to $\theta_{i,k}$ and can thus be discarded. Finally, \eqref{eq:app_ekf3} is obtained since $\Re\{z/j\}=\Im\{z\}$ for $z\in\mathbb C$, and therefore, $p(\bm r_k|\bm\theta_k)$ can be expressed as
	\begin{equation}
		p\left(\bm r_k|\bm\theta_k\right)\approx\gaussian{\bm \theta_k}{\bm{\hat\theta}^{\mathrm f}_{k-1}+\bm{\tilde h}_k}{\mathbf V_k^{-1}},
	\end{equation}
	where $\mathbf V_k$ is defined in \eqref{eq:S} and $\bm{\tilde h}_k\define[\tilde h_{1,k},\dots,\tilde h_{D,k}]^T$ represents the estimated difference between $\bm\theta_k$ and $\bm\theta_{k-1}$, with each of its elements computed as
	\begin{equation}
		\tilde h_{i,k}=\Im\left\{\frac{r_{i,k}e^{-j\hat\theta^{\mathrm f}_{i,k-1}}}{s_{i,k}}\right\}.\label{eq:g}
	\end{equation}
	Finally, using the following identity for the product of two Gaussians \cite[Ch.~A.2]{Rasmussen:2006:GPM:1162254},
	\begin{align}
		\gaussian{\bm x}{\bm a}{\mathbf A}\gaussian{\bm x}{\bm b}{\mathbf B}=\gaussian{\bm b}{\bm a}{\mathbf A+\mathbf B}\gaussian{\bm x}{\bm c}{\mathbf C},
	\end{align}
	where $\mathbf T=(\mathbf I+\mathbf A\mathbf B^{-1})^{-1}$, $\mathbf C=\mathbf T\mathbf A$, and $\bm c=\mathbf T\bm a+\mathbf T\mathbf A\mathbf B^{-1}\mathbf b$,
	allows reducing \eqref{eq:bayes_rec} to
	\begin{align}
		&p(\bm\theta_k|\bm r_1,\dots,\bm r_k)\nonumber\\
		&\approx\gaussian{\bm\theta_k}{\bm{\hat\theta}^{\mathrm f}_{k-1}+\bm{\tilde h}_k}{\mathbf V_k^{-1}}\nonumber\\
		&~~~\cdot\int_{\mathbb R^D}\gaussian{\bm\theta_{k-1}}{\bm\theta_k}{\mathbf Q}\gaussian{\bm\theta_{k-1}}{\bm{\hat\theta}^{\mathrm f}_{k-1}}{\mathbf M^{\mathrm f}_{k-1}}d\bm\theta_{k-1}\nonumber\\
		&=\gaussian{\bm\theta_k}{\bm{\hat\theta}^{\mathrm f}_{k-1}+\bm{\tilde h}_k}{\mathbf V_k^{-1}}\gaussian{\bm\theta_k}{\bm{\hat\theta}^{\mathrm f}_{k-1}}{\mathbf M^{\mathrm f}_{k|k-1}}\label{eq:app_ekf_int}\\
		&\propto\gaussian{\bm\theta_k}{\mathbf T_k\bm{\hat\theta}^{\mathrm f}_{k-1} +\mathbf T_k\mathbf M^{\mathrm f}_{k|k-1}\mathbf V_k\left(\bm{\hat\theta}^{\mathrm f}_{k-1}+\bm{\tilde h}_k\right)}{\mathbf T_k\mathbf M^{\mathrm f}_{k|k-1}}\nonumber\\
		&=\mathcal N_{\bm\theta_k}
		\left(\mathbf T_k\left(\mathbf I_D+\mathbf M^{\mathrm f}_{k|k-1}\mathbf V_k\right)\bm{\hat\theta}^{\mathrm f}_{k-1}\right.\nonumber\\
		&~~~~~~~~~~\left.+\mathbf T_k\mathbf M^{\mathrm f}_{k|k-1}\mathbf V_k\bm{\tilde h}_k,\mathbf T_k\mathbf M^{\mathrm f}_{k|k-1}\right)\nonumber\\
		&=\gaussian{\bm\theta_k}{\bm{\hat\theta}^{\mathrm f}_{k-1}+\mathbf M^{\mathrm f}_{k}\bm h_k}{\mathbf M^{\mathrm f}_{k}},
	\end{align}
	where \eqref{eq:app_ekf_int} is obtained since the integral of a PDF is one, $\mathbf T_k\define(\mathbf I_D+\mathbf M^{\mathrm f}_{k|k-1}\mathbf V_k)^{-1}$, and $\bm h_k\define\mathbf V_k\bm{\tilde h}_k$.
	This leads to the recursive equations in \eqref{eq:ekf_varpred}--\eqref{eq:ekf_corr}.
	
	\section{Derivation of FG messages}
	\label{app:fgmsg}
	Using \eqref{eq:pd_approx} and \eqref{eq:pAPPgauss}, $\Pu{i,k}{s_{i,k}}$ can be computed as
	\begin{align}
		&\Pu{i,k}{s_{i,k}}\propto\int_{\mathbb R^D} p\left(r_{i,k}|s_{i,k},\theta_{i,k}\right)\frac{p(\bm\theta_k|\bm r)}{p\left(r_{i,k}|\theta_{i,k}\right)}d\bm\theta_k\nonumber\\
		&=\int_{\mathbb R}\frac{\cgaussian{r_{i,k}}{s_{i,k}e^{j\theta_{i,k}}}{2\varN{i}}}{p(r_{i,k}|\theta_{i,k})}\left[\int_{\mathbb R^{D-1}}p(\bm\theta_k|\bm r)d\bm{\bar\theta}_{i,k}\right]d\theta_{i,k}\\
		&\approx\int_{\mathbb R}\frac{\cgaussian{r_{i,k}}{s_{i,k}e^{j\theta_{i,k}}}{2\varN{i}}}{\cgaussian{r_{i,k}}{\bar s_{i,k}e^{j\theta_{i,k}}}{2\varNtilde{i}}}\nonumber\\
		&~~~~~~~\cdot\left[\int_{\mathbb R^{D-1}}\gaussian{\bm\theta_k}{\bm{\hat\theta}^{\mathrm s}_{k}}{\mathbf M^{\mathrm s}_{k}}d\bm{\bar\theta}_{i,k}\right]d\theta_{i,k}\\
		&=\int_{\mathbb R}\frac{\cgaussian{r_{i,k}}{s_{i,k}e^{j\theta_{i,k}}}{2\varN{i}}}{\cgaussian{r_{i,k}}{\bar s_{i,k}e^{j\theta_{i,k}}}{2\varNtilde{i}}}\gaussian{\theta_{i,k}}{\hat\theta^{\mathrm s}_{i,k}}{M^{\mathrm s}_{i,k}}d\theta_{i,k}\label{eq:gauss_marg}\\
		&\propto e^{-\frac{|s_{i,k}|^2}{2\varN{i}}}\int_{\mathbb R} 
		\exp\left(\Re\left\{\left(\frac{r_{i,k}s_{i,k}^*}{\varN{i}}-\frac{r_{i,k}\bar s_{i,k}^*}{\varNtilde{i}}\right)e^{-j\theta_{i,k}}\right\}\right)\nonumber\\
		&~~~~~~~~~~~~~~~~~~\cdot\gaussian{\theta_{i,k}}{\hat\theta^{\mathrm s}_{i,k}}{M^{\mathrm s}_{i,k}}d\theta_{i,k}\\
		&\approx e^{-\frac{|s_{i,k}|^2}{2\varN{i}}}\!\!\!\int_{\hat\theta^{\mathrm s}_{i,k}-\pi}^{\hat\theta^{\mathrm s}_{i,k}+\pi}\!\! \exp\left(\!\Re\left\{\!\left(\frac{r_{i,k}s_{i,k}^*}{\varN{i}}-\frac{r_{i,k}\bar s_{i,k}^*}{\varNtilde{i}}\right) e^{-j\theta_{i,k}}\!\right\}\!\right)\nonumber\\
		&~~~~~~~~~~~~~~~~~~~~~~\cdot\tikhonov{\theta_{i,k}}{\frac{e^{j\hat\theta^{\mathrm s}_{i,k}}}{M^{\mathrm s}_{i,k}}}d\theta_{i,k},\label{eq:gauss2tik}
	\end{align}
	where $\tikhonov{z}{\kappa}$ denotes a Tikhonov PDF with a complex parameter $\kappa$ and argument $z$, $\bm{\bar\theta}_{i,k}$ contains all elements of $\bm\theta_k$ except $\theta_{i,k}$, \eqref{eq:gauss_marg} exploits the fact that a multivariate Gaussian is trivially marginalized \cite[Ch.~8]{Rasmussen:2006:GPM:1162254}, and \eqref{eq:gauss2tik} uses the approximation $\gaussian{x}{\mu}{\sigma^2}\approx\tikhonov{x}{e^{j\mu}/\sigma^2}$, which is accurate for small $\sigma^2$ \cite{1504908}. Using the definition of $\xi_{i,k}(s_{i,k})$ in \eqref{eq:xi} gives
	\begin{align}
		&\Pu{i,k}{s_{i,k}}\approx  e^{-\frac{|s_{i,k}|^2}{2\varN{i}}}\int_{\hat\theta^{\mathrm s}_{i,k}-\pi}^{\hat\theta^{\mathrm s}_{i,k}+\pi} \exp\left(\Re\left\{\xi_{i,k}(s_{i,k})e^{-j\theta_{i,k}}\right\}\right)d\theta_{i,k}\nonumber\\
		&\propto  e^{-\frac{|s_{i,k}|^2}{2\varN{i}}}I_0(|\xi_{i,k}(s_{i,k})|)\int_{\hat\theta^{\mathrm s}_{i,k}-\pi}^{\hat\theta^{\mathrm s}_{i,k}+\pi}\tikhonov{\theta_{i,k}}{\xi_{i,k}(s_{i,k})}d\theta_{i,k}\nonumber\\
		&=  e^{-\frac{|s_{i,k}|^2}{2\varN{i}}}I_0(|\xi_{i,k}(s_{i,k})|)\nonumber\\
		&\approx\frac{1}{\sqrt{2\pi|\xi_{i,k}(s_{i,k})|}}\exp\left(|\xi_{i,k}(s_{i,k})|-\frac{|s_{i,k}|^2}{2\varN{i}}\right),\label{eq:finalPu}
	\end{align}
	where $I_0(\cdot)$ is the modified Bessel function of the first kind and zeroth order. From \eqref{eq:finalPu}, \eqref{eq:approxPu}--\eqref{eq:xi} follow.
	
	\section*{Acknowledgment}
	
	The authors would like to express their gratitude to Benjamin J.~Puttnam, Georg Rademacher, and Ruben S.~Lu\'is for providing experimental data that proved to be very beneficial for this work.

\end{document}